\documentclass[a4paper, accepted=2021-06-17]{quantumarticle}

\pdfoutput=1

\usepackage{bm,graphicx}

\usepackage{dsfont}
\usepackage[numbers, sort&compress]{natbib}
\usepackage{amssymb,amsmath,amsfonts}
\usepackage{mathrsfs}
\usepackage{hhline}
\usepackage{epsfig}
\usepackage{epstopdf}
\usepackage[none]{hyphenat}

\usepackage[mathscr]{eucal}
\usepackage{xcolor}
\usepackage{marvosym}  

\usepackage[caption=false]{subfig}

\usepackage[colorlinks]{hyperref}
\hypersetup{
	bookmarksnumbered,
	pdfstartview={FitH},
	citecolor={darkgreen},
	linkcolor={darkred},
	urlcolor={darkblue},
	pdfpagemode={UseOutlines}}
\definecolor{darkgreen}{RGB}{50,190,50}
\definecolor{darkblue}{RGB}{0,0,190}
\definecolor{darkred}{RGB}{238,0,0}

\usepackage[cp1250]{inputenc}
\DeclareInputText{175}{\.Z}     

\newcommand*\xbar[1]{%
   \hbox{%
     \vbox{%
       \hrule height 0.5pt 
       \kern0.5ex
       \hbox{%
         \kern-0.1em
         \ensuremath{#1}%
         \kern-0.1em
       }%
     }%
   }%
} 

\newcommand{\id}{\mathds{1}}

\DeclareMathOperator{\Tr}{Tr}
\DeclareMathOperator{\X}{\mathsf{X}}
\DeclareMathOperator{\Y}{\mathsf{Y}}
\DeclareMathOperator{\Z}{\mathsf{Z}}
\DeclareMathOperator{\diag}{diag}

\newcommand{\ket}[1]{\left|{#1}\right\rangle}
\newcommand{\bra}[1]{\left\langle{#1}\right|}

\newcommand{\oost}{\frac{1}{\sqrt{2}}}

\newcommand{\rmi}{\ensuremath{\mathrm{i}}}

\newcommand{\hd}{\ensuremath{\mathsf{h}}}

\begin{document}

\title{The shape of higher-dimensional state space: Bloch-ball analog for a qutrit}

\author{Christopher Eltschka}
\affiliation{Institut f\"ur Theoretische Physik, Universit\"at Regensburg, D-93040 Regensburg, Germany}

\author{Marcus Huber}
\affiliation{Atominstitut,  Technische  Universit{\"a}t  Wien,  1020  Vienna,  Austria}
\affiliation{Institute for Quantum Optics and Quantum Information - IQOQI Vienna, Austrian Academy of Sciences, Boltzmanngasse 3, 1090 Vienna, Austria}

\author{Simon Morelli}
\affiliation{Institute for Quantum Optics and Quantum Information - IQOQI Vienna, Austrian Academy of Sciences, Boltzmanngasse 3, 1090 Vienna, Austria}
\affiliation{Atominstitut,  Technische  Universit{\"a}t  Wien,  1020  Vienna,  Austria}

\author{Jens Siewert}
\email{jens.siewert@ehu.eus}
\affiliation{Departamento de Qu\'{i}mica F\'{i}sica, Universidad del Pa\'{i}s Vasco UPV/EHU, E-48080 Bilbao, Spain}
\affiliation{IKERBASQUE Basque Foundation for Science, E-48009 Bilbao, Spain
}

\date{22 June, 2021}

\begin{abstract}
Geometric intuition is a crucial tool to obtain deeper insight into many 
concepts of physics. A paradigmatic example of its power is the Bloch ball, the geometrical representation for the state space of the simplest possible quantum system, a two-level system (or qubit).
However, already for a three-level system (qutrit) the state space has eight dimensions, so that its complexity exceeds the grasp of our three-dimensional space of experience.
This is unfortunate, given that the geometric object describing the state space of a qutrit has a much richer structure and is in many ways more representative for a general quantum system than a qubit.
In this work we demonstrate that, based on the Bloch representation of quantum states, it is possible to construct a three dimensional model for the qutrit state space that captures most of the essential geometric features of the latter.
Besides being of indisputable theoretical value, this opens the door to a new type of representation, thus extending our geometric intuition beyond the simplest quantum systems.
\end{abstract}

\maketitle

\section{Introduction}

Nowadays virtually every student of quantum mechanics learns
about the Bloch sphere and the Bloch ball as the geometrical 
representations of pure and mixed states of qubits, respectively.
These objects have become indispensable for developing an intuition 
of elementary
concepts such as basic quantum operations and 
the action of decoherence~\cite{Nielsen_2000},
or more advanced topics like the Majorana representation of 
symmetric multi-qubit states~\cite{Martin_2010}.
Nonetheless, the more experienced practitioner in the field of quantum mechanics
is aware of various shortcomings of the Bloch ball if systems of higher
dimension $d>2$ are to be discussed. For example, the entire surface
of the Bloch ball is covered by pure states, whereas most of the boundary of
higher-dimensional state spaces is formed by mixed states. Those parts of
the boundary may either be flat or curved, however, their curvature is
different from that of the pure-state surface parts, which 
actually represents a single unitary orbit.
Another important fact not shown 
by the Bloch ball is that not every orthogonal transformation can be applied
to any quantum state: Whenever a state is not an element of the inscribed
sphere of the state space of maximum radius there 
exist rotations in Bloch space that take it outside the state space and
therefore are not allowed.
Consequently, it is a long-standing question whether it is possible to
construct a consistent
three-dimensional model for higher-dimensional state spaces 
$\mathcal{Q}_d$, $d\geq 3$, 
that captures at least a part of these important geometric features.

\section{Bloch ball for qubits}
%
Let us briefly recapitulate the properties of the Bloch ball for qubits,
$d=2$. According to Fano~\cite{Fano1954,Fano1957} density operators 
of qubits are parametrized by using the Pauli
matrices $\sigma_x$, $\sigma_y$, $\sigma_z$ and the identity $\id_2$
\begin{subequations}
\begin{align}
     \rho\  & =\ \frac{1}{2}\left( s_0 \id_2+x\ \sigma_x + y\ \sigma_y
                                         +z\ \sigma_z \right)
\\
     s\ & =\ \Tr\left(\sigma_s\rho\right)\ \ ,
\end{align}
\label{eq:bloch2}
\end{subequations}
with real numbers $s\in\{x, y, z\}$, $|s|\leq 1$,
and the normalization $s_0=1$.
The state space 
$\mathcal{Q}_2$ of all qubit density operators is represented by a ball
of radius 1 about the origin of $\mathbb{R}^3$, that is, each point 
$(x,y,z)$ of this ball corresponds to exactly one state $\rho$.
Pure states lie on the surface (forming a connected set), 
whereas mixed states are
located inside the Bloch sphere, with the fully mixed state at the center.
Interestingly, the convex combinations of two states, 
$\rho=\lambda \rho_1+(1-\lambda)\rho_2$
($0\leq \lambda\leq 1$),
are given by the straight line connecting the points of $\rho_1$ and $\rho_2$.
A common choice of computational basis states is $\{\ket{0},\ket{1}\}$, which correspond to
north and south pole of the sphere and form the regular simplex
$\Delta_1$, the special case of $\Delta_{d-1}$ for $d=2$. This shows that 
the vectors in $\mathbb{R}^3$ belonging to the basis states are {\em 
not} orthogonal.
These vectors are orthogonal in $\mathbb{R}^4$ including the direction of $s_0$, but their projections into $\mathbb{R}^3$ lose this property.

Every density matrix can be obtained by unitarily rotating a diagonal density matrix, that resembles a classical probability distribution. Since for $d=2$ the special unitary group SU(2) is the universal cover group of the rotation group SO(3), every rotation of the Bloch ball $Q_2\subset\mathbb{R}^3$ has a corresponding unitary rotation in the state space. From this perspective, the Bloch ball is obtained by all possible rotations of the simplex $\Delta_1$ in $\mathbb{R}^3$.

To the best of our knowledge, the idea that the 
parametrization in Eq.~\eqref{eq:bloch2} entails a useful visualization for the 
non-unitary dynamics of a 
spin $\frac{1}{2}$, e.g., in a
situation of radiation damping
was put forward by Feynman and co-workers~\cite{Feynman1957}.

\section{A Bloch-ball analog for a qutrit}
%
Over the years much work has been done to elucidate the geometry of the state space of higher level systems, with special focus on the qutrit state space $\mathcal{Q}_3$~\cite{Bengtsson_2006,Bloore_1976,Ramachandran_1979,Byrd_2003,Kimura_2003,Kimura_2005,Mendas_2006,Boya_2008,Rau_2009,Dunkl_2011,Kurzynski_2009,Sarbicki_2012,Bengtsson_2012,Tabia_2013,Goyal_2016,Kurzynski_2016,Weis_2018,Xie_2020,Paraoanu_2020,Sharma_2021}.
To develop an intuition for the full high-dimensional geometry of the qutrit state space, subsets, cross sections and projections onto two and three dimensions were extensively studied~\cite{Bengtsson_2006,Bloore_1976,Kimura_2003,Mendas_2006,Dunkl_2011,Bengtsson_2012,Sarbicki_2012,Goyal_2016,Kurzynski_2016,Weis_2018,Xie_2020} and multi-parameter representations of qutrit states were developed~\cite{Rau_2009,Kurzynski_2016,Braun_2020,Paraoanu_2020,Sharma_2021}.
While these approaches can reproduce many geometric properties correctly, they do not give a global view of the Bloch body.

In this section we review known facts about the higher dimensional state space, with focus on dimension $d=3$, collecting a list of requirements we wish our model to reproduce. We then construct a three dimensional global model of the state space, that reproduces astonishingly many properties of $\mathcal{Q}_3$.


\subsection{Qutrit geometry}
In analogy to qubits, density matrices describing the state of a $d$-level quantum system can be parametrized by the identity $\mathds{1}_d$ and $d^2-1$ traceless Hermitian matrices~\cite{Fano1957,Bengtsson_2006,Bertlmann_2008}. So the quantum state space $\mathcal{Q}_d$ can be represented by a subset in $\mathbb{R}^{d^2-1}$, constrained by inequalities that arise from the positivity of the density operators~\cite{Kimura_2003, Kimura_2005}.
While it is still true that the quantum state space $\mathcal{Q}_d$ is obtained by rotating the simplex $\Delta_{d-1}$ in $\mathbb{R}^{d^2-1}$, it is no longer a sphere as not all rotations are allowed. In fact, for $d>2$ the special unitary group SU($d$) is a proper subgroup of the rotation group SO($d^2$-1), meaning that the quantum state space $\mathcal{Q}_d$ is a proper subset of the $d^2-1$ dimensional (Hilbert-Schmidt) ball.

For qutrits the density matrices are
parametrized by the identity $\id_3$ and the normalized Gell-Mann matrices
$\X_j$, $\Y_k$ ($j,k=1,2,3$), and $\Z_1$, $\Z_2$ (see Appendix),
\begin{subequations}
\begin{align}
     \rho\ = &\ \frac{1}{3}\left( \id_3+ \sum_{j=1}^3 x_j \X_j+ 
                     \sum_{k=1}^3 y_k \Y_k+ \sum_{l=1}^2 z_l \Z_l
                         \right)
\label{eq:bloch3a}
\\
     x_j = & \Tr\left(\X_j\rho\right)\, , \ 
     y_k =  \Tr\left(\Y_k\rho\right)\, ,\  
     z_l =  \Tr\left(\Z_l\rho\right)\, ,  
\label{eq:bloch3b}
\end{align}
\label{eq:bloch3}
\end{subequations}
with real numbers $x_j$, $y_k$, and $z_l$, hence their space has
eight dimensions. A Euclidean metric, corresponding to that of 
our everyday geometric experience, is induced in this space by
the Hilbert-Schmidt norm,
\mbox{$\|\rho_1-\rho_2\|_2$}$\equiv\sqrt{3\Tr\left[(\rho_1-\rho_2)^2\right]}$.

The geometric properties of $\mathcal{Q}_3$ we wish for a global
model to reproduce were thoroughly discussed by 
Bengtsson et al.\ in Ref.~\cite{Bengtsson_2012}.
\\
i) Most importantly, $\mathcal{Q}_3$ is a convex set with the topology of
a ball, so the model should share this characteristics. There are no
pieces of lower dimension attached to it (``no hair'' condition). 
\\
ii) The actual Bloch body is neither a polytope nor a smooth object.
\\
iii) The Bloch body has an outer sphere of radius $\sqrt{2}$ and an inner sphere
of radius $1/\sqrt{2}$\ \cite{note}.
\\
iv) The pure states (rank 1) form a 
connected set on the surface at maximum distance $\sqrt{2}$
from the completely mixed
state $\frac{1}{3}\id_3$. 
Its measure is zero compared to that of $\mathcal{Q}_3$.
In particular we aim at prominently displaying the three pure states 
corresponding to the preferred basis for the model.
\\
v) Density matrices on the surface of $\mathcal{Q}_3$ are of rank 1 or 2,
whereas states inside $\mathcal{Q}_3$ are of full rank (rank 3).
\\
Finally, there are some additional properties specifically
related to the nature of $\mathcal{Q}_3$ as a convex set:
\\
vi) The set of quantum states is self-dual.
\\
vii) All cross sections of $\mathcal{Q}_3$ do not have non-exposed faces.
 All corners of two-dimensional projections of $\mathcal{Q}_3$ are
polyhedral.


\subsection{A three dimensional model for a qutrit}
Surprisingly, it is indeed possible to find an object in $\mathbb{R}^3$
representing $\mathcal{Q}_3$ that obeys most
of the requirements in this list. In fact, there are (at least) two solutions
with slightly different advantages. First we construct
an object that fulfills properties i--iv) and also partially v). 
It represents a valid model for $\mathcal{Q}_3$ that we call $Q_3^{(1)}$.
Then we show that this first solution can
be extended to a model $Q_3^{(2)}$ that possesses in particular also 
the property vi). Yet that model does not obey vii). 

It is well known that the
computational basis states in $d$ dimensions form the corners
of a regular simplex
$\Delta_{d-1}$ in $\mathcal{Q}_d$~\cite{Bengtsson_2006,Bengtsson_2012}, that is, for a qutrit a basis will
be represented by $\Delta_2$, an equilateral triangle. We insist
that our model for $\mathcal{Q}_3$ displays and emphasizes one
particular basis $\{\ket{0},\ket{1},\ket{2}\}$, 
because the mapping between physical states and points in the model
will depend on this choice.
Hence, we use the coordinates
[cf.~Eq.~\eqref{eq:bloch3b}] 
\begin{align*}
     z_1\ =\ \Tr\left(\Z_1 \rho\right)\ \ ,\ \ \ \
     z_2\ =\ \Tr\left(\Z_2 \rho\right)\ \ ,
\end{align*}
to {\em faithfully} represent
the diagonal matrices as a simplex $\Delta_2$ in the horizontal 
coordinate plane. In particular, the basis state $\ket{2}$ corresponds
to $(z_1,z_2)=(0,-\sqrt{2})$, whereas the states $\ket{0}$ and $\ket{1}$
are located in $(\pm\sqrt{\frac{3}{2}},\sqrt{\frac{1}{2}})$, respectively.
The completely mixed state
lies at the origin $(0,0)$.

\begin{figure}[t]
  \centering
  \includegraphics[width=.995\linewidth]{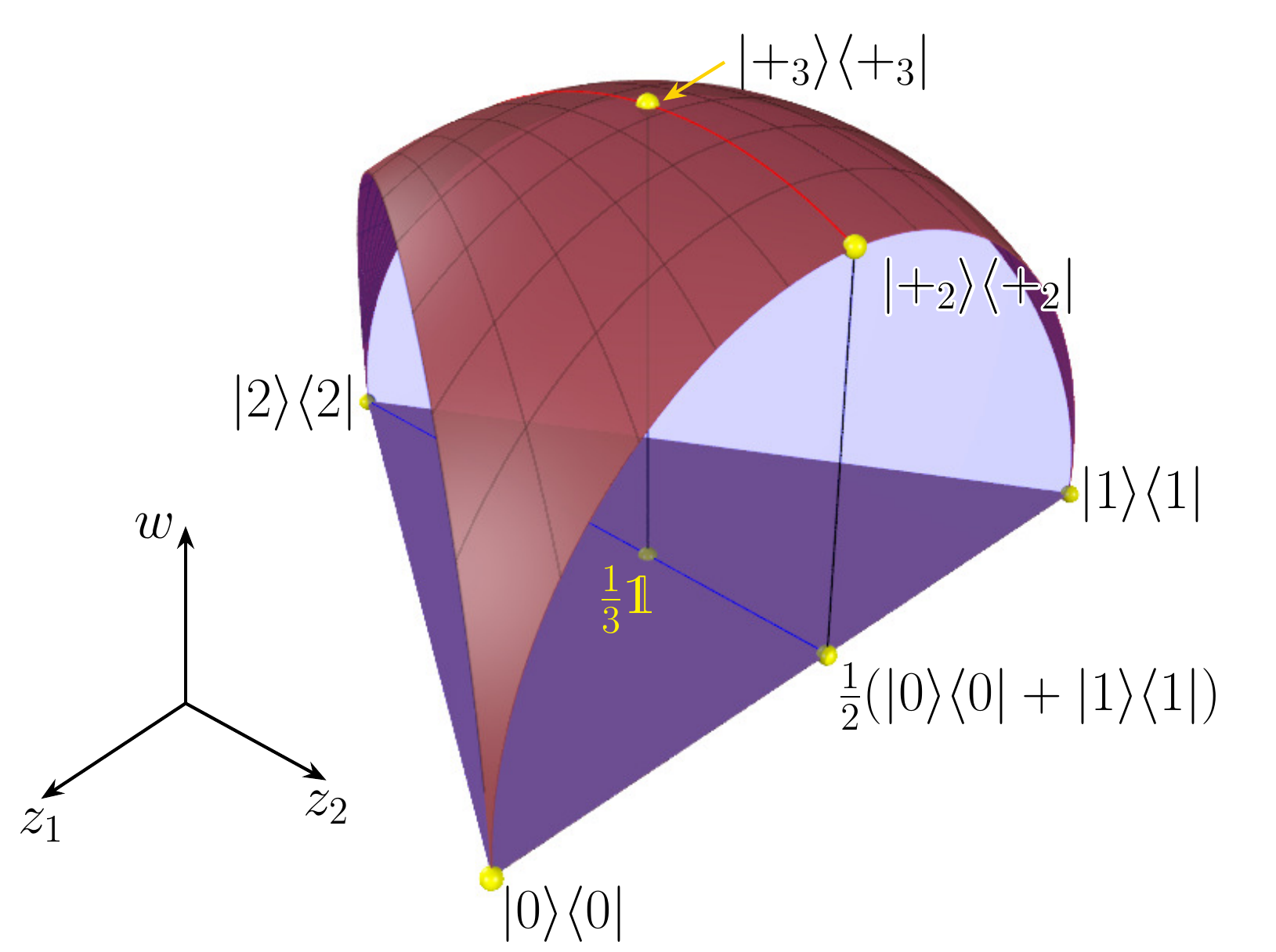}
  \caption{The model $Q_3^{(1)}$ of the qutrit Bloch body
           $\mathcal{Q}_3$
           according to Eq.~\eqref{eq:Q31}. The location of
           the basis states $\ket{j}$ is specified. The semicircular
           surface in the foreground is the image of the Bloch ball
           spanned by the states $\ket{0}$ and $\ket{1}$ and, in this
           representation, has the ``north pole'' 
           $\ket{+_2}=\oost(\ket{0}+\ket{1})$. The point with the
           largest $w$ coordinate is the image of the maximally
           coherent superposition $\ket{+_3}=\frac{1}{\sqrt{3}}
           (\ket{0}+\ket{1}+\ket{2})$.
    Note that all the pure states with $|\langle j | \psi \rangle|^2=\frac{1}{3}$ get mapped to this point, in particular all the bases which are mutually unbiased with $\{\ket{0},\ket{1},\ket{2}\}$.}
  \label{fig:fig1}
\end{figure}

For the remaining six coordinates we are left with only one direction in
$\mathbb{R}^3$. Here, we propose to use the coordinate
\begin{align}
             w\ = &\ \sqrt{\sum_{j=1}^3 x_j^2+y_j^2}\ \ ,
\label{eq:vert}
\end{align}
which assumes only non-negative values. 
Hence we have our first model for $\mathcal{Q}_3$, 
\begin{align}
   Q_3^{(1)} = \{ (z_1,z_2,w)  & \in\mathbb{R}^3\ \mathrm{s.t.} \
        \nonumber\\
                & \mathrm{Eqs.}~\eqref{eq:bloch3b},\eqref{eq:vert}
                             \ \forall \rho\in\mathcal{Q}_3
                      \} .
\label{eq:Q31}
\end{align}
The interpretation for
the coordinates of a point $P=(z_1,z_2,w)$
is simple.
Imagine the state $\rho$ corresponding to $P$ written as a 
sum of its diagonal and offdiagonal parts, 
\[
\rho\ =\ \rho_{\mathrm{diag}}\ +\ \rho_{\mathrm{offdiag}}\ \ .
\]
While the distance of $P$ from the origin in the plane equals the 
Hilbert-Schmidt length $\sqrt{3\Tr(\rho_{\mathrm{diag}}^2)-1}$
of the diagonal part of the Bloch vector for $\rho$,
the vertical distance of $P$ from the plane is the Hilbert-Schmidt length
of this Bloch vector's offdiagonal part, 
$\sqrt{3\Tr(\rho_{\mathrm{offdiag}}^2)}$. The result is shown in Fig.~\ref{fig:fig1}.
%
%

\subsection{Global geometric properties}
Let us analyze the properties of this first model for $\mathcal{Q}_3$
with respect to our list of requirements. The surface of this object
corresponds to a hemisphere of radius $\sqrt{2}$, where the three spherical
segments beyond the triangle in the plane are cut off. Evidently, it is
both convex and simply connected, without anything else attached to it.
The flat surfaces of the cuts are connected with the smooth upper boundary by a sharp corner, so the object is neither smooth nor a polytope.
Since our model preserves the length of the Bloch vector, it is still circumscribed by
an outer sphere of radius $\sqrt{2}$ and center at the origin.
The radius of the inner (hemi-)sphere coincides
with that of the in-circle of the simplex $\Delta_2$ in the horizontal
plane and equals $1/\sqrt{2}$.

The spherical part of the surface above the ground plane corresponds
to the set of pure states, hence to density matrices of rank 1.
They form a simply connected surface of measure zero at maximal distance $\sqrt{2}$ from the origin.
The cuts are half circles, in fact, these 
flat surfaces are the images of the three two-dimensional Bloch balls
corresponding to the pairs of states $\{\ket{0},\ket{1}\}$,
$\{\ket{0},\ket{2}\}$, and $\{\ket{1},\ket{2}\}$ (cf.~also Fig.~\ref{fig:fig1}). This means,
these surfaces correspond to states of at most rank 2. 
It is understood that the base of $Q_3^{(1)}$ represents
an artificial cut -- similar to the base of the sculpture of a bust -- 
that does not represent a boundary of $\mathcal{Q}_3$.
The states on the other surfaces are of lower rank, all states of rank 3 reside in the interior of the model.
However rank-2 states that have all of the basis vectors $\ket{0}$, $\ket{1}$, $\ket{2}$ 
in their span are located
{\em inside} $Q_3^{(1)}$, although in $\mathcal{Q}_3$ they
are part of the boundary.
But the rank-2 states do not cover the entire interior of $Q_3^{(1)}$.
Evidently, all points inside the inner sphere are of rank 3, since rank-2 states have purity
of at least $\frac{1}{2}$. However the set of points corresponding only to rank-3 states
is larger than that; see Sec.~\ref{sect:localprop} for details.
Conversely, the interior points outside that set correspond both to rank-2 and rank-3 states.
In particular, in agreement with property v), the images of rank-3 states
cover the complete interior.
Thus, we have established that our construction $Q_3^{(1)}$ has all the desired
properties i--iv) of our list, property v) is
at least partially satisfied.

We may ask how the coordinates in $Q_3^{(1)}$ are related to those 
of the actual state space $\mathcal{Q}_3$.
Clearly our model is neither a projection nor a cross section. Instead, it 
resembles a representation in cylindrical coordinates: Our ``diagonal 
coordinates'' $z_1$, $z_2$ correspond to two longitudinal axes, whereas
each $r_j=\sqrt{x_j^2+y_j^2}$ (where $j=1,2,3$) may be viewed as a 
radial coordinate belonging to mutually orthogonal directions. 
Our model does not display the polar angles and shows only the
total radial distance.


\subsection{Local algebraic properties}
\label{sect:localprop}
We have mentioned that the simplex $\Delta_2$ represents the diagonal
states faithfully, it is isomorphic to that set just as the
Bloch ball is isomorphic to $\mathcal{Q}_2$.
But our three-dimensional model $Q_3^{(1)}$ cannot represent all states of the eight-dimensional state space faithfully.
In fact, a general point of $Q_3^{(1)}$ corresponds to infinitely many states and the simplex representing the basis states $\ket{j}\!\bra{j}$, $j=0,1,2$ is the only set of states with a unique preimage.
States mapped onto the same point in the model belong to the equivalence class of states with the same diagonal entries and purity, they form a five-dimensional manifold in the original state space $\mathcal{Q}_3$. These subspaces are closed under the action of unitary operators that commute with $\Z_1$ and $\Z_2$, in particular diagonal unitary operators. As a consequence the model is invariant under these transformations.

\begin{figure}[tb!]
  \centering
  \includegraphics[width=.995\linewidth]{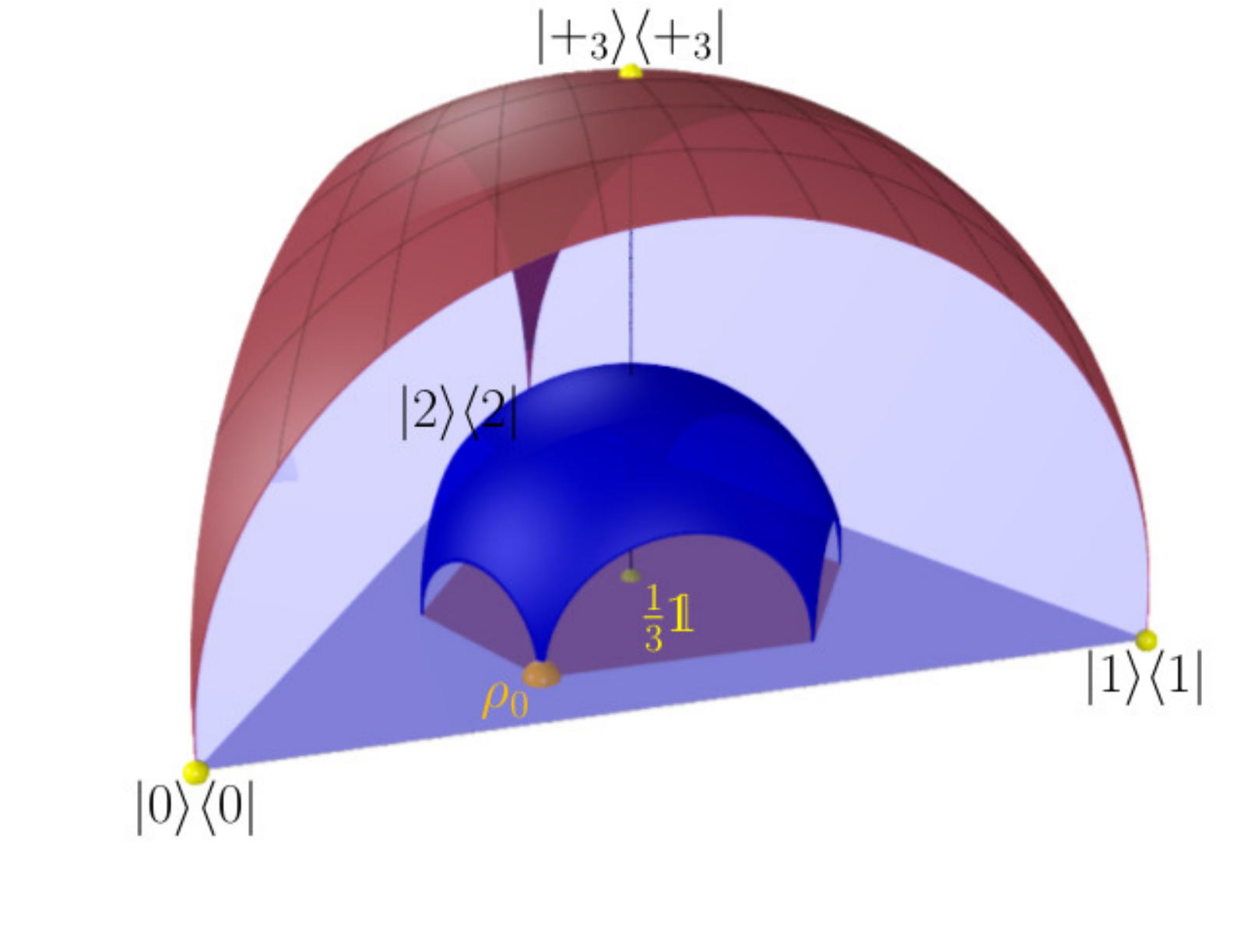}
  \caption{The SU(3) orbit of the state $\rho_0=\frac{6}{10}\ket{0}\!\bra{0}+\frac{3}{10}\ket{1}\!\bra{1}+\frac{1}{10}\ket{2}\!\bra{2}$ is represented by the blue spherical surface. The image of $\rho_0$  together with the other five points in the $z_1$-$z_2$ plane forms the Birkhoff polytope (see text). The eigenvalue vectors of the diagonal states inside the polytope are majorized by that of $\rho_0$, $(\frac{6}{10},\frac{3}{10},\frac{1}{10})$.
    }
  \label{fig:unitary}
\end{figure}

The action of general unitary transformations, however, is displayed by our model. Since the purity remains unchanged, all points in the unitary orbit have the same distance from the origin. This orbit forms a subset of a sphere, its shape depends on the eigenvalues of the transformed state. As previously elucidated, a point in the interior represents a whole subspace of states with possibly different eigenvalues, so it makes little sense to talk about the orbit of a point in our model. However, diagonal states in the ground triangle are depicted faithfully and therefore uniquely identify a unitary orbit. One only needs to use the eigenbasis of the state to investigate which points can be reached by unitary transformations. The permutations of the diagonal entries correspond to SU(3) rotations, therefore the unitary orbit includes six points on the triangle (for non-degenerate eigenvalues). Since the eigenvalues of a matrix majorize the diagonal entries, the orbit remains on a sphere above this hexagon. Vice versa, every point on the sphere above the hexagon can be reached through a unitary transformation, see Fig.~\ref{fig:unitary}. If the eigenvalues are degenerate, the hexagon becomes a triangle. A special case are the pure states, the unitary orbit of a pure state coincides with the upper surface of $Q_3^{(1)}$.

This visualization establishes a direct connection to the action of doubly stochastic matrices on classical probability distributions. A classical probability distribution $\mathsf{p}$ majorizes exactly the probability distributions $M.\mathsf{p}$, where $M$ is a doubly stochastic matrix. The set of distributions majorized by $\mathsf{p}$ is called the Birkhoff polytope~\cite{Bengtsson_2006}. The action of doubly stochastic matrices on classical probability distributions corresponds to the action of unitary transformations on normalized diagonal matrices. That is, for every doubly stochastic matrix $M$ there exists a unitary matrix $U$, such that $M.\mathsf{p}=\diag(U.\mathsf{D}.U^\dagger)$, where $\mathsf{D}$ is a diagonal matrix with $\mathsf{p}=\diag(\mathsf{D})$. In the model this is visualized by unitarily transforming a diagonal state and then projecting it onto the ground triangle. This exactly reproduces the Birkhoff polytope, but it also generalizes it in the sense that the norm of the offdiagonal part remains visible.
%
%

The images of SU(3) orbits also allow us to  identify the set of images of
rank-2 states. The rank-2 diagonal states are exactly the sides of the base triangle.
Therefore the rank-2 states are given by all the orbits of those points. It is readily seen
that the boundary arcs connecting two triangle sides form half-circular cones that
each have one of the basis states as apex, and are tangential to the insphere at their base.
The points inside those cones, as well as the points inside the inner sphere, are not covered
by those orbits, and therefore correspond only to states of rank 3.
This rank-3 only region is convex; indeed it is the convex hull of the interior
of the insphere and the interior of the base triangle.

Note that the mirror image of this region and its boundary forms the lower part of our
second model $Q_3^{(2)}$ which will be described in Sec.~\ref{sect:selfdual}.

The map that takes the qutrit state space to our  model is decidedly nonlinear, but amazingly
retains some linear properties of the state space for the image.
Any convex combination of two diagonal qutrit states
$\delta=\lambda \delta_1+(1-\lambda)\delta_2$ with $\lambda\in(0,1)$
is again a diagonal state and its image lies on the straight line connecting the image points of
$\delta_1$ and $\delta_2$.
But even the convex combinations $\rho=\lambda \sigma +(1-\lambda)\delta$ of an arbitrary qutrit state $\sigma$ with 
a diagonal state $\delta$ are located on the straight line connecting $\sigma$ and $\delta$.
The mixture of the diagonal parts is faithfully represented, but in this case, this holds also for the offdiagonal parts, because $\delta_{\mathrm{offdiag}}$ simply vanishes.
Consequently, for any problem involving two arbitrary states we can find a visualization of their mixture
by means of a straight line: It is enough to consider the problem in the eigenbasis of one of the states.

Another noteworthy aspect of our model $Q_3^{(1)}$ is the separate
treatment of diagonal and offdiagonal parts of the state. The diagonal 
matrices may be viewed as states in a classical probability 
space~\cite{Bengtsson_2012}. They become nonclassical by adding coherences, i.e., an offdiagonal part. The $w$ coordinate
in our model is the 2-norm for the offdiagonal part of the state.
As the 1-norm of the offdiagonal part is established as a coherence
measure~\cite{Baumgratz_2014} 
and the 1-norm is lower-bounded by the 2-norm, our model explicitly
shows a lower bound to the coherence of a given quantum state.
Correspondingly, the maximally coherent state 
$\ket{+_3}=\frac{1}{\sqrt{3}}(\ket{0}+\ket{1}+\ket{2})$ is located at the
north pole of the hemisphere.

\section{Self-dual qutrit geometry}
\label{sect:selfdual}
%
The model we have discussed so far does not meet the requirements
v) and vi) of our list. In particular it is not self-dual, that is,
our three-dimensional Bloch vectors do not fulfill the following
condition. Let $\vec{\xi}$ be the Bloch vector belonging to a state on the 
boundary of the state space. Then for the set of Bloch vectors $\vec{\eta}$
defining the dual hyperplanes enveloping the state space on 
the side opposite to $\vec{\xi}$  we have~\cite{Bengtsson_2012}
\begin{align}
   \vec{\xi}\cdot\vec{\eta}   \ =\  -1
\ \ .
\end{align}
Our model so far is not self-dual simply because for vectors $\vec{\xi}$
on the surface with $w>0$ the dual planes do not touch the lower 
boundary of our Bloch body model. We will now demonstrate
that the model can be extended by adding a ``lower part'' with
coordinates $w<0$ so that the entire object becomes self-dual.

\subsection{Self-dual extension of $Q_3^{(1)}$}
%
To achieve self-duality, consider first the pure states $\Pi$ with 
$\Tr(\Pi^2)=\Tr(\Pi)=1$. 
If we denote the vector of Gell-Mann matrices
by $\vec{\hd}$ and the Bloch vector of the state $\Pi$ by $\vec{\pi}$,
we can write $\Pi=\frac{1}{3}(\id_3+\vec{\pi}\cdot\vec{\hd})$. 
Recall that reversing the sign of all coordinates in the qubit Bloch sphere
amounts to a point reflection operation at the maximally mixed state.
This fact suggests that we might try reversing the sign of $w$ in a process
of reversing all of the coordinate signs, for example in an operation 
of the kind $\rho\ \longrightarrow\ \alpha \id_3 -\rho$ (where $\alpha>0$).
We have to make sure that the result is again positive.
In Ref.~\cite{Rungta_2001} the so-called universal state inversion 
map $\mathcal{S}(\rho)=\nu(\id-\rho)$ was introduced; it guarantees
that the result actually is a state. We choose the
prefactor $\nu=\frac{1}{2}$ so as to normalize the resulting 
qutrit state and write
\begin{align}
    \mathcal{S}(\rho)\ \equiv\ \tilde{\rho}\ =\ \frac{1}{2}(\id_3-\rho)
\ \ .
\label{eq:state-inv}
\end{align}
The inverted state $\tilde{\rho}$ is positive and for pure states
we have $\Tr(\Pi\tilde{\Pi})=0$. Therefore,
\begin{align}
     0\ =\ 1\ +\ \vec{\pi}\cdot \vec{\tilde{\pi}}\ \ ,
\label{eq:dual-pure}
\end{align}
that is, the Bloch vector $\vec{\tilde{\pi}}$ defines the dual plane for
$\vec{\pi}$ (and vice versa). As the image of $\vec{\pi}$ determines a 
point in the spherical part of the boundary of our model, also the
image of $\vec{\tilde{\pi}}$ lies on a spherical surface, however,
with half the radius because of the prefactor $\nu=\frac{1}{2}$. The result
is that the spherical part of our model $Q^{(1)}_3$ gets replicated 
in the region
$w<0$, whereat it is point-reflected at the origin and scaled down
by a factor $\frac{1}{2}$.
%
\begin{figure}[t!]
  \centering
  \includegraphics[width=.99\linewidth]{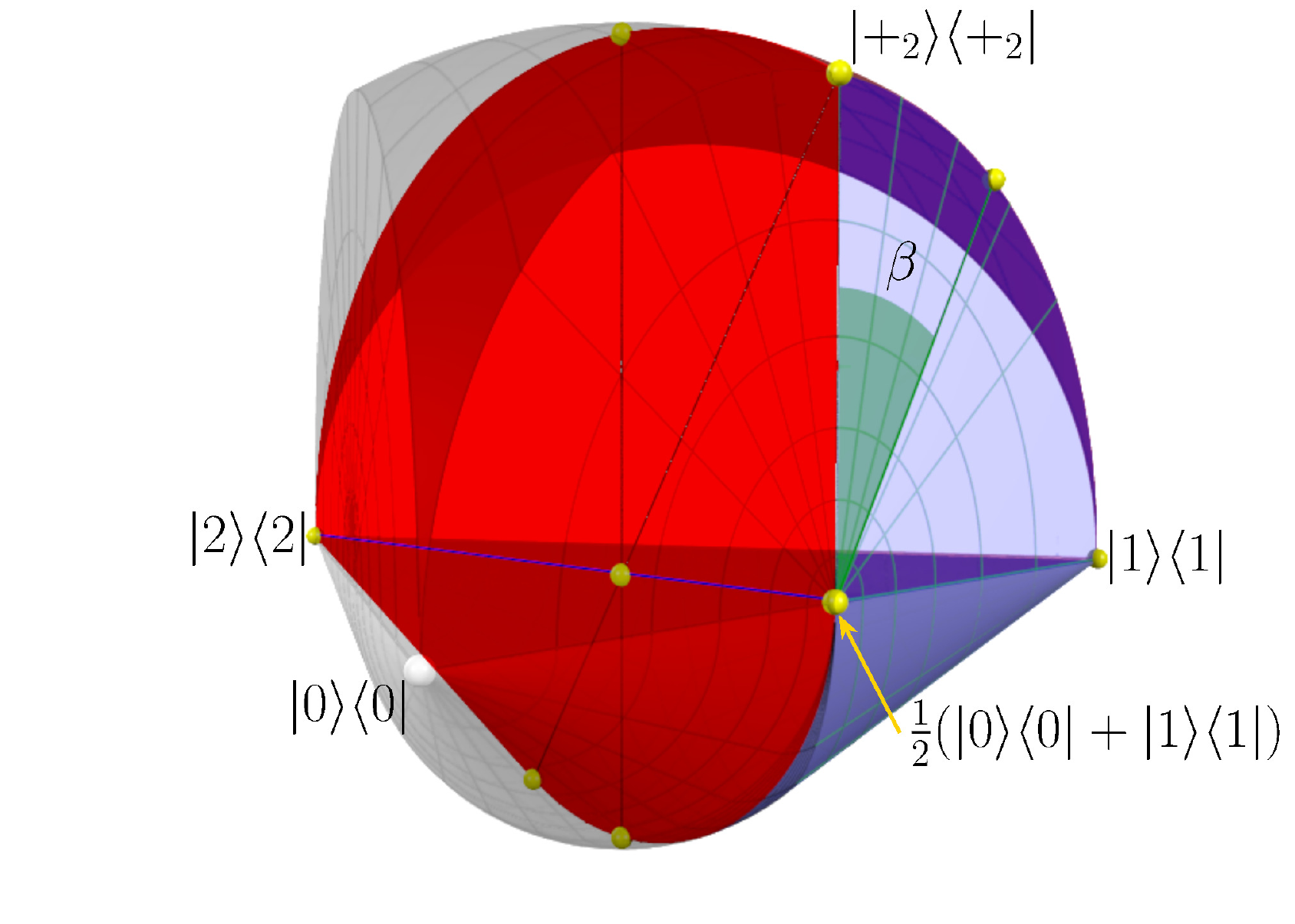}
  \caption{Constructing the dual of the surface of the model
           $Q_3^{(1)}$. 
           Red: Cross section for determining the dual states 
           $\tilde{\rho}(p)$ of the mixtures $\rho(p)$ in Eq.~\eqref{eq:rhop}.
           Gray: We have included the part that has been cut from 
           the full Bloch body (see Fig.~\ref{fig:fig4}, which shows
           the complete object $Q_3^{(2)}$ from the same point of view). By rotating the section highlighted in red by the angle $\beta$ (see text) one obtains the lower conical part of $Q^{(2)}_3$.
    }
  \label{fig:fig3}
\end{figure}
%
%

This reasoning cannot be applied for the mixed boundary states of the
model. 
Rather, we apply state inversion $x_j\rightarrow -x_j$, 
$y_k\rightarrow -y_k$, $z_l\rightarrow -z_l$ and determine the boundary
states by explicit calculation. Consider, for example, the mixtures (cf.~Fig.~\ref{fig:fig3})
\begin{align}
\rho(p)\ =\ p\ket{+_2}\!\bra{+_2}\ +\ \frac{1-p}{2}(\ket{0}\!\bra{0}+\ket{1}\!\bra{1})
\label{eq:rhop}
\end{align}
with $0\leq p\leq 1$. It is straightforward to show that the dual states
are given by
\begin{align}
  \tilde{\rho}(p)\ =\ &\frac{1}{3}\bigg( \id_3 - \frac{\sqrt{2}}{1+3p} \Z_2
                                              - \frac{\sqrt{6}\ p}{1+3p} \X_1
                                 \bigg)
\nonumber\\
     =\  \frac{1}{3}\bigg( \id_3 & - 
         (1-q)\sqrt{2} \Z_2 - q\left[ 
                  \frac{\sqrt{2}}{4} \Z_2 +\frac{\sqrt{6}\ p}{4} \X_1
                                \right]
                                 \bigg)
\label{eq:rhop-dual}
\end{align}
with $q=\frac{4p}{1+3p}$.
That is, also $\tilde{\rho}(p)$ lies on a straight line connecting
$\ket{2}\!\bra{2}$ and  the dual of $\ket{+_2}\!\bra{+_2}$.
An analogous calculation
applies to all mixtures of 
$\frac{1}{2}(\ket{0}\!\bra{0}+\ket{1}\!\bra{1})$ 
with pure states on the circumference of 
the $\{\ket{0},\ket{1}\}$ ``Bloch sphere''.
Instead of considering the cross section in the $z_2$-$w$ plane 
for the states $\tilde{\rho}(p)$ we can rotate this plane about the $z_2$
axis by an angle $\beta$ (see Fig.~\ref{fig:fig3}), 
so that the pure state on the circumference
becomes $\cos{\left(\frac{\pi-2\beta}{4}\right)}\ket{1}+
\sin{\left(\frac{\pi-2\beta}{4}\right)}\ket{0}$. 

\begin{figure}[t!]
  \centering
  \includegraphics[width=.99\linewidth]{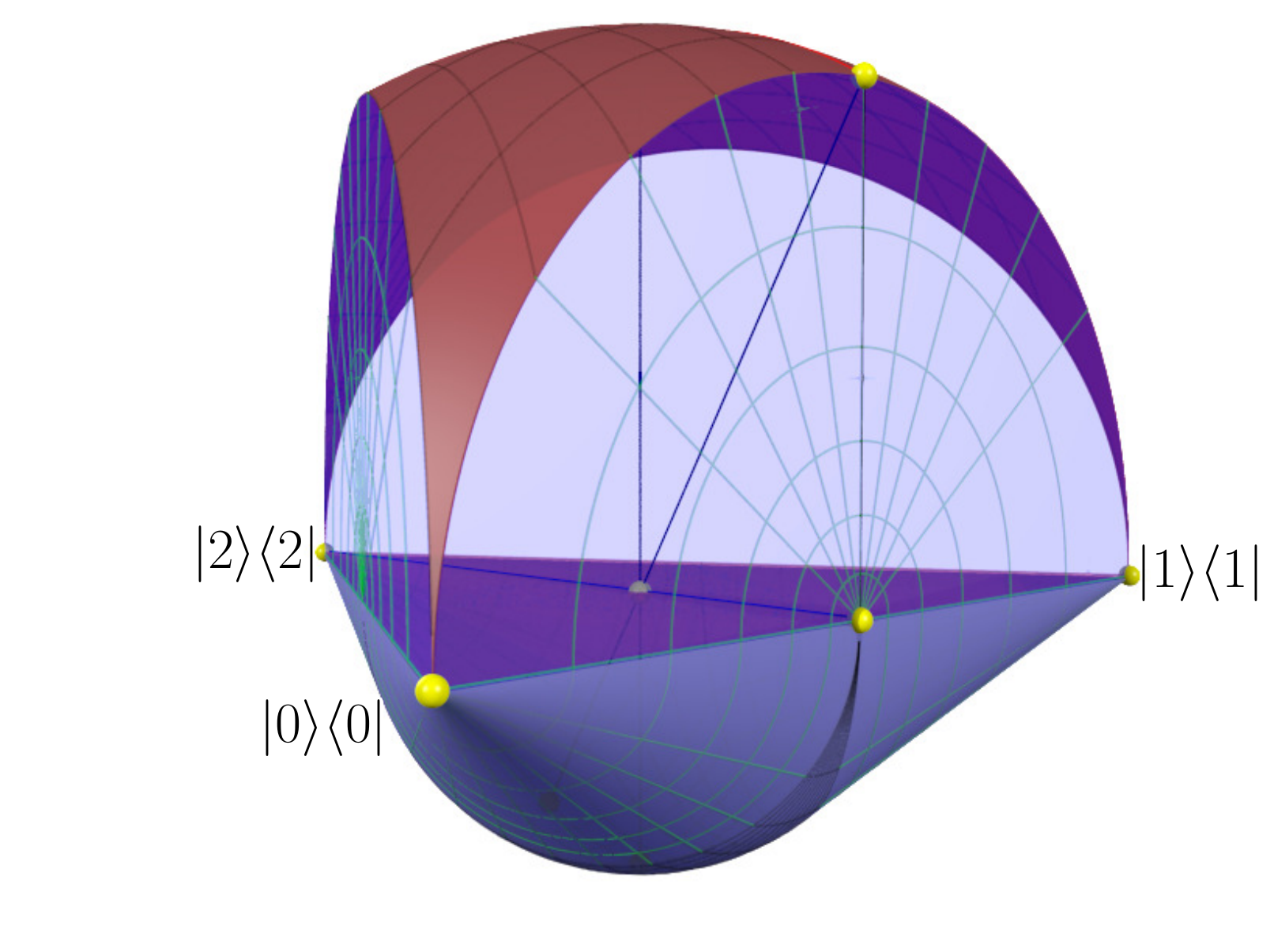}
  \caption{The self-dual model $Q_3^{(2)}$ of the qutrit state space $\mathcal{Q}_3$.
    }
  \label{fig:fig4}
\end{figure}

The shape of the corresponding 
parts of the cross section remains the same as in the one shown in Fig.~\ref{fig:fig3}.
Consequently, the dual
states of the qubit Bloch ball $\{\ket{0},\ket{1}\}$ are located on a
half circular cone with apex in $\ket{2}\!\bra{2}$ (cf.~Fig.~\ref{fig:fig4}). 
Due to the three-fold symmetry of  $Q_3^{(1)}$ the regions near the other
pure states $\ket{0}\!\bra{0}$, $\ket{1}\!\bra{1}$ 
can be constructed analogously.

Once we have found the shape [i.e., the boundary
coordinates $w_{\mathrm{max}}(z_1,z_2)$ and $w_{\mathrm{min}}(z_1,z_2)$]
of the self-dual object,
we also need to provide a rule to decide whether a given state 
$\rho$ from inside $\mathcal{Q}_3$ is mapped to the 
``upper'' ($w>0$) or the ``lower'' ($w<0$)  part of the object.
In other words, from $\rho$ we can determine the coordinates
$z_1$, $z_2$,  $|w|$ as well as $w_{\mathrm{min}}(z_1,z_2)$,
but how can we decide about the sign of $w$?

In the following we describe a procedure to define
this mapping $\rho\ \longrightarrow\ P(z_1,z_2,w)$. 
It is both straightforward and consistent (however, there might be
other choices).
Let $\rho$ be an arbitrary qutrit state with smallest eigenvalue
$\lambda_{\mathrm{min}}$. First we subtract as much of the fully
mixed state so that one eigenvalue vanishes, i.e.,
\begin{equation}
   \rho^{\prime}\ =\ \frac{1}{1-3\lambda_{\mathrm{min}}}
                 \left(\rho - \lambda_{\mathrm{min}}\id_3
                                              \right)
\end{equation}
is a rank-2 state with Bloch coordinates $(z_1^{\prime},z_2^{\prime},
|w^{\prime}|)$. Then
\begin{equation}
\ \left.  \begin{array}{l}
         |w'|> |w_{\mathrm{min}}(z_1',z_2')|  \\*[2.0mm]
         |w'|\leq |w_{\mathrm{min}}(z_1',z_2')| 
                  \end{array}
          \right\}\ 
                    \Longrightarrow\ 
             \left\{  \begin{array}{l}
                      w>0  \\*[2mm]
                       w<0
                      \end{array}
             \right.\ \ .
\label{eq:sign-w}
\end{equation}
This implies that in the case of
equality the state $\rho'$ is located at the lower boundary in $Q_3^{(2)}$.
Moreover, together with $\rho'$ also all its mixtures with $\frac{1}{3}\id_3$
belong to the lower part of $Q_3^{(2)}$, ensuring its compactness.
The meaning of this definition is clear: In comparison to the structure
of $Q_3^{(1)}$, in $Q_3^{(2)}$ we redistribute the 
location of the rank-2 states $\rho'$.
If the purity of $\rho'$ is large enough for its diagonal coordinates
$(z_1',z_2')$ it needs to remain in the upper part, 
$w>0$, otherwise it goes to the lower part, $w<0$.

This concludes the construction of the self-dual model $Q^{(2)}_3$ for the 
qutrit Bloch body, see Fig.~\ref{fig:fig4}:
\begin{align}
    Q_3^{(2)} = \big\{(z_1,  & z_2,w) \in \mathbb{R}^3\ \mathrm{s.t.}\
    \nonumber\\
            &    \mathrm{Eqs.}~\eqref{eq:bloch3b},
                             \eqref{eq:vert},\eqref{eq:sign-w}
                             \forall \rho\in\mathcal{Q}_3 \big\}\  .
\end{align}
We note that the extension to a self-dual object can be applied
as well to the half-circle model $Q_2^{(1)}$ of the $d=2$ Boch ball.
Then, however, the rule given in Eq.~\eqref{eq:sign-w}
to determine the sign of $w$ (without any addendum) does not improve the model.
This is because in $d=2$ all offdiagonal elements
can be made real and positive by applying a single appropriate 
diagonal unitary to the 
state. In contrast, for $d>2$ 
at most $2(d-1)$ offdiagonal matrix elements 
of a generic state can simultaneously be made positive
by applying a diagonal unitary.

%
%
\subsection{More convexity properties and relation to previous work}
%
Convexity properties of the state space in the context of the Bloch-ball representation for $d=2$ (and even for $d=3$) were discussed early on, e.g., by Bloore~\cite{Bloore_1976} who analyzed the stratification of the state space with respect to the matrix rank. For a qutrit, all states of reduced rank $r<3$ are part of the boundary. The four-dimensional set of rank-1 states $\ket{\psi}\!\bra{\psi}$ forms the extremal states of the convex state space, as they do not have a convex decomposition into other states
$\rho_1$ and $\rho_2$, 
$\ket{\psi}\!\bra{\psi}\neq \lambda \rho_1+(1-\lambda)\rho_2$ with 
\mbox{$0\leq \lambda\leq1$}.
Geometrically they are part of the outer sphere, because they have distance $\sqrt{2}$ from the origin. 

As described in the previous section, the lower part of our self-dual model $Q_3^{(2)}$ consists of three half-circular cones,
each with a basis state as their apex, which end at a half-sized point mirror image of the
pure-state spherical section, that is, at the insphere. This description matches exactly the
one of the internal boundary separating the points corresponding
to rank-2 states from those corresponding only to rank-3 states in the model
$Q_3^{(1)}$, except that it is below the base triangle rather than above.
Also in the previous section, it was established that rank-2 states are mapped to the lower half exactly if doing so does not locate them outside the
model.

This allows us to determine the image of the set of rank-2 states in $Q_3^{(2)}$.
The only rank-2 states that fit into the lower part are exactly those which in
$Q_3^{(1)}$ lie on the internal boundary between rank-2 states and rank-3 only states. Therefore all rank-2 states that get mapped into the lower half of the self-dual model are
mapped onto the model's surface, while the interior of the lower part consists only of
rank-3 states, just as criterion v) requires.

However, the same is not true in the upper part. Almost all the rank-2 states in the interior of ${Q}_3^{(1)}$ are not moved to the lower part and therefore are also in the interior
of ${Q}_3^{(2)}$.
Only the points on the internal rank boundary in $Q_3^{(1)}$ are moved to the boundary of
$Q_3^{(2)}$, leaving only rank-3 states on the internal boundary.
That is, criterion v) is still not perfectly satisfied. However, the new model $Q_3^{(2)}$ is self-dual by construction and hence obeys the criterion vi). 

The last criterion vii) is 
critical because also the self-dual model $Q_3^{(2)}$ cannot reproduce
these properties. This can be observed in the highlighted 
cross section in Fig.~\ref{fig:fig3}.  The point corresponding to 
$\frac{1}{2}(\ket{0}\!\bra{0}+\ket{1}\!\bra{1})$ is non-exposed and
counts as a non-exposed face of 
the two-dimensional cross section~\cite{Bengtsson_2012}.
As these non-exposed faces occur at points where the self-dual extension is attached to $Q_3^{(1)}$ one might speculate that their occurrence is a price to pay for having such an extension.
Moreover, the highlighted cross section in Fig.~\ref{fig:fig3} coincides with the projection of $Q_3^{(2)}$
in the direction of the \mbox{$z_1$ axis}. Neither the corner corresponding
to $\ket{2}\!\bra{2}$ nor that of $\ket{+_2}\!\bra{+_2}$ is polyhedral.

The self-duality of $Q_3^{(2)}$ is particularly interesting, because it enables us to analyze how findings from previous work are featured in our model. As a first example we consider the duality of boundary states on outer and inner spheres. As was elucidated in Refs.~\cite{Kimura_2005,Goyal_2016},
boundary states on the outer sphere have their dual counterparts on the inner sphere, and vice versa. This property can be observed in Figs.~\ref{fig:fig3} and~\ref{fig:fig4} and is guaranteed by construction via the universal state inversion, Eq.~\eqref{eq:state-inv},~\eqref{eq:dual-pure}.

It is noteworthy that the universal state inversion (or reduction~\cite{Horodecki_1999}) map finds an explicit geometric application here. Until now it was mainly associated with entanglement properties of multi-party states, although the relevance of geometrical concepts was implicit also in that context (cf., e.g.,~\cite{Eltschka_2018}).

One of the most notable qutrit three-sections is the so-called {\em obese tetrahedron} (or three-dimensional elliptope). It was already noted by Bloore~\cite{Bloore_1976} and thoroughly studied by Goyal and co-workers~\cite{Goyal_2016} (see Sec.~5.4 and Fig.~5 in Ref.~\cite{Goyal_2016}). Its barycenter is the completely mixed state and the corners are given by the four non-orthogonal states $\frac{1}{\sqrt{3}}(\ket{0}\pm\ket{1}\pm\ket{2})$.
The faces are bulgy, because the tetrahedron contains with its corner points also their duals on the opposite side. The image of the tetrahedron in $Q_3^{(2)}$ is the part of the vertical axis belonging to the model: The corner points are all mapped to the north pole, the barycenter to the origin and the centers of the faces to the point with $z_1=z_2=0$ and $w=-\frac{1}{\sqrt{2}}$.

The last example we mention here is the conical three-section, Sec.~5.1 and Fig.~2 in Ref.~\cite{Goyal_2016}. Instead of the [128] section
we consider the (unitarily equivalent) [138] section
which indeed has a three-dimensional image in  $Q_3^{(2)}$.
The states of the cone are formed by all convex combinations of the real states of the $\{\ket{0},\ket{1}\}$
Bloch ball (a circular disk) and the state $\ket{2}\!\bra{2}$.
Therefore, one might expect the image to include for $w>0$  the corresponding half-cone with the semicircular surface at $z_2=\frac{1}{\sqrt{2}}$ and the apex at $z_2=-\sqrt{2}$. However, this is only
partially correct, because a part of the half-cone gets mapped to values $w<0$ according to our rule, Eq.~\eqref{eq:sign-w}. This includes the complete half-cone attached to $\ket{2}\!\bra{2}$ for values $w<0$  and $z_2\leq -\frac{\sqrt{2}}{4}$. Moreover, also the half-cone with apex in the origin and base at $z_2=-\frac{\sqrt{2}}{4}$ with base radius $\frac{\sqrt{6}}{4}$ gets mapped to $w<0$. The union of both parts at $w<0$ and $w\geq 0$ corresponds to the half-cone mentioned before, that is, to one half of the actual conic three-section.

\section{Quantum channels}

\begin{figure*}[t!]
  \centering
  \includegraphics[width=.7\paperwidth]{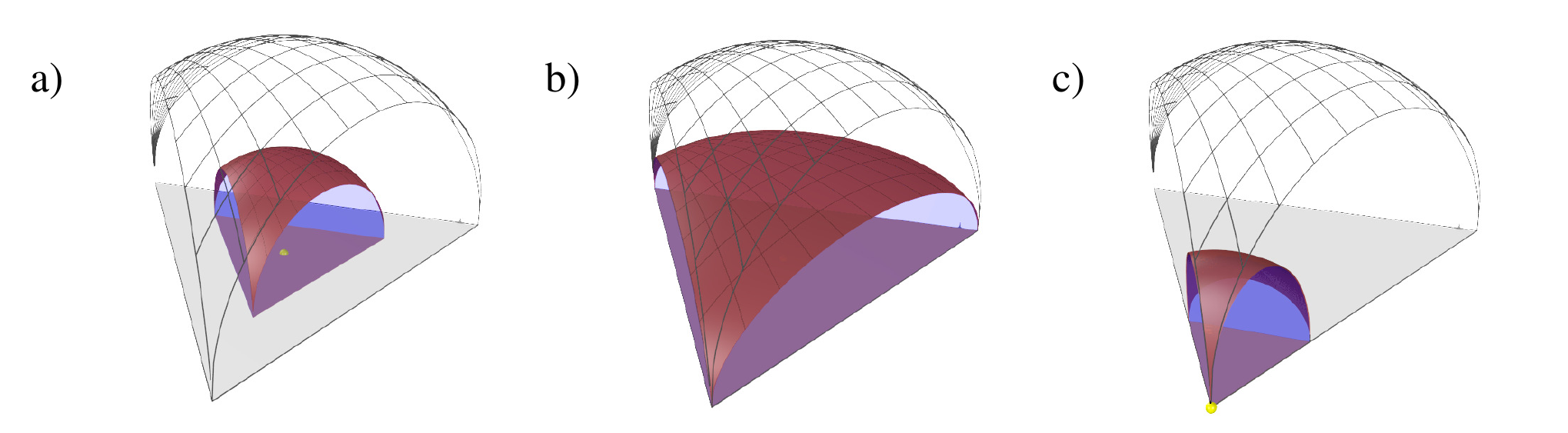}
  \caption{Action of quantum channels on the qutrit state space in $Q^{(1)}_3$ representation: a)
           depolarizing, b) phase-damping, and c)
           amplitude-damping channel.
    }
  \label{fig:fig5}
\end{figure*}

In order to give yet another demonstration that our method directly connects
to the well-known properties of the Bloch sphere/ball visualization
we show the action of three standard decohering channels
(cf.~Ref.~\cite{Nielsen_2000}) on a qutrit. 
The depolarizing channel corresponds to driving a state towards 
the completely mixed state $\frac{1}{3}\id_3$, 
whereas phase-damping amounts to mixing
a state with its diagonal part. Amplitude-damping describes the
relaxation to one of the basis states. Following the
discussion of $Q^{(1)}_3$  the action of these channels
on the qutrit Bloch body is evident and completely analogous
to what is known for the $d=2$ Bloch ball (cf.~Fig.~\ref{fig:fig5}).

The channels are characterized by their Kraus operators $K_j$ that
obey the relation $\sum_j K_j^{\dagger}K_j=\id_3$. For the depolarizing
channel we have $K_0=\sqrt{1-8\gamma/9}\ \id_3$,
$K_j=\sqrt{\gamma/9}\  \hd_j$, where $\ \hd_j$ $(j=1\ldots 8)$
stands as a shortcut for all the Gell-Mann matrices, so that
\begin{align}
    \rho'(\gamma)\ =\ (1-\gamma)\ \rho\ +\ \gamma\ \id_3\ \ .
\end{align}
The parameter $\gamma\in [0,1]$ describes the strength of the channel action.

The Kraus operators for phase damping (or pure dephasing) in its simplest version are  $K_0=\sqrt{1-2\gamma/3}\ \id_3$,
$K_1=\sqrt{\gamma/3}\ \Z_1$, $K_2=\sqrt{\gamma/3}\ \Z_2$.

Under the action of the depolarizing and the phase-damping channel, 
the shrinking of the Bloch body
model $Q_3^{(1)}$ occurs proportionally along straight lines, and
the images of states that are represented by the same point will again
be represented by the same point in the shrunken model. The same is,
however, not true for the amplitude-damping channel.

The amplitude damping channel considered is of the form given by
Grassl et al.~\cite{Grassl_2018}, with $\gamma_{01}=\gamma_{02}=:\gamma$ and
$\gamma_{12}=0$, so that the states $\lvert 1\rangle$ and $\lvert 2\rangle$ are
damped equally. 
That is, 
\begin{align}
K_0\ =\ & \ket{0}\!\bra{0}+\sqrt{1-\gamma}\ket{1}\!\bra{1}
                        +\sqrt{1-\gamma}\ket{2}\!\bra{2}
\nonumber\\
K_1\ =\ & \sqrt{\gamma}\ket{0}\!\bra{1}
\\
K_2\ =\ & \sqrt{\gamma}\ket{0}\!\bra{2}
\nonumber \ \ .
\end{align}
%
%
%
This results in non-equal scaling of the non-diagonal elements.
Whereas $\rho_{01}$ and $\rho_{02}$ scale with a prefactor of
$\sqrt{1-\gamma}$, the element $\rho_{12}$ gets a prefactor 
of $(1-\gamma)$.

This does not matter for diagonal states ($w=0$), which are
mapped into the model uniquely, and which are therefore again scaled
linearly, by
\begin{align}
z_1 &\mapsto z_1' =
\sqrt{\frac{3}{2}}+\left(z_1-\sqrt{\frac{3}{2}}\right)(1-\gamma)\\
z_2 &\mapsto z_2' =\frac{1}{\sqrt{2}} + \left(z_2 - \frac{1}{\sqrt{2}}\right)(1-\gamma)
\end{align}
Note that these two coordinates are scaled the same way for all the states.

For pure states, the absolute value of the offdiagonal elements is
uniquely determined by the diagonal elements, and since only the
absolute value enters into the $w$ coordinate, again each point of the
original model gets mapped to only one point of the shrunken model. In
particular, the $w$ coordinate of pure states gets mapped according to

\begin{widetext}
\begin{align}
  w'^2 = (1  -\gamma )^2  & \left[\frac{2}{3}
            \left(1-z_2^2-\frac{z_2}{\sqrt{2}}\right)
    \  + \ z_1 \left(\frac{2}{\sqrt{3}}
   z_2-\sqrt{\frac{2}{3}}\right)\right]
\nonumber \\
     &\ +\ (1-\gamma )\ \left[\frac{4}{3}\ -\ z_1^2
   +\left(\sqrt{\frac{2}{3}}-
 \frac{2}{\sqrt{3}} z_2\right)
 z_1-\frac{z_2^2}{3}+\frac{\sqrt{2}z_2}{3}
\right]\ \ .
\end{align}
\end{widetext}
For non-diagonal mixed states, states with different distribution of
absolute values in the offdiagonal matrix elements get mapped to the
same point in the model. Therefore states that are represented by the
same point will get mapped by the channel to states represented by
different points. It is obvious that the value of $w$ will always be
mapped somewhere into the range $[(1-\gamma)^2w,(1-\gamma)w)$, but that range
will not be exhausted for all source points.

\section{Conclusion}
%
We have presented two ways of visualizing the eight-dimensional state
space of a qutrit in $\mathbb{R}^3$, thereby preserving a number
of essential geometric properties of $\mathcal{Q}_3$,
cf.~Ref.~\cite{Bengtsson_2012}. 
Depending on the context one or the other representation 
may appear more useful.  
Our findings are relevant, because the properties of $Q^{(1)}_3$ and
$Q^{(2)}_3$ are much closer to those of arbitrary higher-dimensional
state space than the Bloch ball for qubits. 

Thus we hope our results are helpful to develop a more precise intuition for  objects in higher dimensions, which are commonly encountered in quantum  information science. 

It is conceivable that there are more possibilities for such visualizations along the lines of this work. The essential features of this visualization persist even if we use it to represent state spaces of 
dimensions \mbox{$d>3$} (splitting the generalized Gell-Mann into a diagonal and off-diagonal part in an identical fashion). Finding the most useful geometrical representation of those and categorizing them is a direction that we believe to yield further fruitful insight in the future.

\acknowledgments
This work was funded by the German Research Foundation Project
EL710/2-1 (C.E., J.S.), by grant PGC2018-101355-B-100 (MCIU/FEDER,UE)
and Basque Government grant IT986-16 (J.S.).
C.E.\ and J.S.\ acknowledge Klaus Richter's support for this project. M.H. and S.M. acknowledge funding from the Austrian Science Fund (FWF) through the START project Y879-N27 and the stand-alone project P 31339-N27.

\section{Appendix}
%
For completeness, we provide the definition for the 
Gell-Mann matrices that we use in Eqs.~\eqref{eq:bloch3a}
and \eqref{eq:rhop-dual}.
Please note that we use a normalization to the dimension $d=3$
(as opposed to the normalization to 2 conventionally
used in high-energy physics). Moreover, also the enumeration differs
from the usual one and is adapted to our coordinate description:
\begin{align*}
\X_1\ =\ & \sqrt{\frac{3}{2}}
           \left[ \begin{array}{ccc}
                    0 & 1 & 0\\
                    1 & 0 & 0\\
                    0 & 0 & 0
                \end{array}
           \right]
\ \  & 
\Y_1\ =\ & \sqrt{\frac{3}{2}}
           \left[ \begin{array}{ccc}
                    0 & -\rmi & 0\\
                    \rmi & 0 & 0\\
                    0 & 0 & 0
                \end{array}
           \right]
\\[2mm]
\X_2\ =\ & \sqrt{\frac{3}{2}}
           \left[ \begin{array}{ccc}
                    0 & 0 & 1\\
                    0 & 0 & 0\\
                    1 & 0 & 0
                \end{array}
           \right]
\ \  & 
\Y_2\ =\ & \sqrt{\frac{3}{2}}
           \left[ \begin{array}{ccc}
                    0 & 0 & -\rmi \\
                    0 & 0 & 0\\
                    \rmi & 0 & 0
                \end{array}
           \right]
\\[2mm]
\X_3\ =\ & \sqrt{\frac{3}{2}}
           \left[ \begin{array}{ccc}
                    0 & 0 & 0\\
                    0 & 0 & 1\\
                    0 & 1 & 0
                \end{array}
           \right]
\ \  & 
\Y_3\ =\ & \sqrt{\frac{3}{2}}
           \left[ \begin{array}{ccc}
                    0 & 0 & 0 \\
                    0 & 0 & -\rmi\\
                    0 & \rmi & 0
                \end{array}
           \right]
\\[2mm]
\Z_1\ =\ & \sqrt{\frac{3}{2}}
           \left[ \begin{array}{ccc}
                    1 & 0 & 0\\
                    0 & -1 & 0\\
                    0 & 0 & 0
                \end{array}
           \right]
\ \  & 
\Z_2\ =\ & \frac{1}{\sqrt{2}}
           \left[ \begin{array}{ccc}
                    1 & 0 & 0 \\
                    0 & 1 & 0\\
                    0 & 0 & -2
                \end{array}
           \right]\ \ .
\end{align*}
\bibliographystyle{apsrev4-1fixed_with_article_titles_full_names}
\bibliography{bibfile.bib}

\end{document}